\documentclass[%
 reprint,
 amsmath,amssymb,
 aps,
]{revtex4-2}

\usepackage{graphicx}
\usepackage{dcolumn}
\usepackage{bm}

\begin{document}

\preprint{APS/123-QED}

\title{Calibrating Coupling for Collaboration with Kuramoto}%

\author{Alexander C. Kalloniatis}
\email{alexander.kalloniatis@defence.gov.au}
\author{Timothy McLennan-Smith}%
 \email{timothy.mclennan-smith@defence.gov.au}
\affiliation{%
Defence Science and Technology Group,
Canberra, Australia
}%

\date{\today}

\begin{abstract}
We calibrate a Kuramoto model inspired representation of peer-to-peer
collaboration using data on the maximum team-size where coordination breaks down. The Kuramoto model is modified,
normalising the coupling by
the degree of input and output nodes,
reflecting dispersion of cognitive
resources in both absorbing incoming- 
and tracking outgoing-information. We find
a critical point, with 
loss of synchronisation as the
number of nodes grows and analytically determine this
point and calibrate the coupling 
with the known maximum team size. We test that against the known `span of control' for a leader/supervisor organisation. Our results suggest larger maximum team sizes than early management science proposes, but are consistent with recent studies.

\end{abstract}

\maketitle


\section{\label{sec:level1}
Introduction}

Data from the International Software Benchmarking Standards Group (ISBSG)  \cite{rodriguez2012empirical}, indicates that 'Projects with an average team size of 9 or more people ... are less productive ...'. Data on span of control in health systems  \cite{cathcart2004span}
show that employee engagement drops from the 16-20 band onwards.
In this paper we develop a physics-based model to provide
insight into such phenomena.
The extension of physics
to understand such collective human
behaviour is a burgeoning field
\cite{BALL20021}, much using
statistical physics for insight into large aggregated human groups
\cite{Castellano2009}.
However, modelling human
group interactivity as dynamical systems 
\cite{HORST2010158,Agbanusi2018} has potential
for smaller organisations. We adapt the Kuramoto model of
synchronising oscillators \cite{kuramoto2003chemical}
for such dynamics, showing how
data on the largest
size of a viable team may calibrate such models, and be
applied to other organisational designs.

The Kuramoto model for an unweighted
network is written as
\begin{eqnarray}
\dot{\theta}_i = \omega_i
- \sum_{j=1}^N
\sigma_{ij} A_{ij} \sin(\theta_i-\theta_j),
\label{KurEq}
\end{eqnarray}
where $\theta_i$ is the phase of
the i-th oscillator,
$\omega_i$ its intrinsic (or native)
frequency drawn from 
some distribution $g(\omega)$, $\sigma_{ij}$ a matrix of
coupling strengths between
individuals, and $A_{ij}$
the unweighted adjacency matrix
of a graph (with element one for connected
nodes and zero otherwise).
Kuramoto \cite{kuramoto2003chemical}
originally proposed
Eq.(\ref{KurEq}) for a complete graph
($A_{ij}=1$ for $i\neq j$) and
uniform coupling normalised by
the degree of each node,
$\sigma_{ij}=\sigma/d_i$ where the degree is given by $d_i = (N-1)$.
For $N\rightarrow \infty$ he was
able to compute analytically
the critical coupling
\begin{eqnarray}
\sigma_c = \frac{2}{\pi g(0)}
\label{KurCoupl}
\end{eqnarray}
for symmetric unimodal 
probability distribution
for frequencies, $g(\omega)$, centred on
$\omega=0$. The model has been examined for
numerous graphs,
regular (trees \cite{DekTay2013} and rings \cite{Rogge_2004})
and random (uniform \cite{ichinomiya2004frequency}, scale free
\cite{oh2007synchronization}, and small world
\cite{hong2002synchronization}).

This much reviewed model \cite{acebron2005kuramoto,arenas2008synchronization,dorogovtsev2008critical,dorfler2014synchronization} applies across scientific fields: 
robotic systems \cite{Mizumoto2010},
ecosystems \cite{vandermeer2021new} and various social phenomena such as rhythmic audience-clapping \cite{Neda2000}, scientific collaboration \cite{Pluchino2006}
and rhythmic crowd-walking \cite{Strogatz2005}.
Encouraged by this, we have applied the Kuramoto model
to human organisations. A stochastic version of the model may represent
organisational structures and processes
in a military headquarters \cite{kalloniatis2020EJOR}. 
The model may be extended to incorporate
competition over resources
\cite{ahern2021}
and tensions across multiple
organisations \cite{McLennan-Smith2021,Zuparic2022}.

This mapping of Eq.(\ref{KurEq})
to an organisation
is straightforward.
The phase is the state
of an individual's cognitive 'perception-action' cycle
\cite{Neisser1967}
or `Observe-Orient-Decide-Act' loop
\cite{boyd1987organic}.
Frequencies are individual decision-making speeds.
The adjacency matrix 
represents communications through the organisational structure
and business processes (both formal and informal). The coupling
is the strength of
those interactions,
where `coupling' is a well-known
though heretofore qualitatively
described human organisational concept \cite{weick1976educational,perrow2011normal,HollenSpitz2012}. With frequencies
interpreted as decision-making {\it speeds} and phases as perception-action {\it states}, our organisational instantiation of
the Kuramoto model
differs from that in \cite{dekker2007studying}.
We note that such cyclic 
cognitive models
transcend the human
species \cite{freeman2002limbic,schoner1998action}.

Previously we calibrated the coupling by comparing multiple scenarios for the same organisation \cite{kalloniatis2020EJOR}.
However, in this we overlooked known natural limits on
organisational sizes.
We show how, using these
bounds on viable -- namely
{\it synchronisable} -- organisations, the coupling may be fixed and validated.

The coupling normalisation is key, already recognised
for taking the large $N$ limit
\cite{gomez2007synchronizability}. Previously, we assumed a linear dissipation of coupling by node degree $d_i$ and
$\sigma_{ij}=\sigma/ d_i$ reflecting the individual's management of concurrent relationships. This is incorrect for human teams: Kuramoto's 
result Eq.~\eqref{KurCoupl} with degree normalisation gives a coupling where synchronisation breaks down independent
of $N$ for large $N$.
This implies that finite cognitive effort enables synchronisation across an arbitrary 
number of team-partners, manifestly
wrong. A stronger dissipation of coupling
is required to give
a natural maximum size of a team defined through a critical value of $N$ at which team synchronisation breaks down.

We provide an intuitive basis for the appropriate normalisation, explore this numerically for
different forms of frequency 
distributions, and analytically determine the
critical $N$.
There nevertheless remains a free scale, either the
frequency-distribution width
or the maximum frequency.
We calibrate the 
ratio of the coupling to frequency
scale, 
and cross-check against similar
empirical estimates of
the number of subordinates
a supervisor is capable of managing, so-called 'span of control' in organisational design.
The literature is less unified
with recommendations here, though
we find a result consistent with
those for a larger span
of control.
We conclude with 
prospects for future 
cross-disciplinary research 
across human, physical and technology sciences.

\section{Teams, Relationships and Cognition}
A team is typically defined as a group of
people working 
interdependently toward
a common goal or purpose \cite{Dyer1984TeamRA}.
They are `complex adaptive systems' through
their internal workings and interactions
with their environment \cite{IlgenHollenbeck2005}.
Successful teams
share a mental model
of the work to be done and of their own
roles and states within that purpose
\cite{IlgenHollenbeck2005,BakerDaySalas2006}.
This provides for an implicit mechanism
of coordination. For us the
important quality is the cognitive
load on team individuals
in maintaining that 'task/team state' mental model. 
Intuitively, the cognitive
load follows the number of interactions within a flat all-to-all coupled team,
$N(N-1)/2$ for $N$ members. 
There comes a threshold where a human becomes incapable of
managing multiple concurrent
working relationships for a given task, triggering break-down in
coordination \cite{Steiner1972}.
The above ISBSG recommendation exemplifies such 
organisational heuristics and other related ones:
the Brooks effect \cite{Brooks1975}, where adding additional
staff to a late software project only makes it later; and
`social loafing' or the Ringelmann effect
\cite{INGHAM1974371}, where
beyond a certain point addition of new members
sees individuals contributing decreasing amounts
of effort. 

Beyond task-directed contexts,
Dunbar's number is the maximum number of individuals
with which one is capable of maintaining
{\it social} relationships, also linked
to human cognitive capacities under biological evolution \cite{dunbar_1993}.
Whereas Dunbar's number is high --
estimated at 150 -- for 
task-based relationships the
viable limit is lower, varying amongst authors between five and
eight team members. Recent research puts this even higher \cite{MaoWatts2016}, though with
semi-realistic tasks these
studies use an
experimental digital system
with simulated artefacts
enabling memory storage, extending individual mental capacities. Such technology enables coordination within larger
teams \cite{Groth1999}.
In this respect,
our paper focuses on `pure' teams collaborating without technological
support systems interfaces. Finally here we note that
whereas limits of viable team sizes seem
independent of the types of tasks that
are performed, we do not expect this to apply
to the time-scale of decision-making.
This leaves an open scale in the model.

\section{Normalisation and Numerical exploration}
As discussed,
the naive linear choice of degree normalisation yields an $N-$independent critical coupling. 
We argue a quadratic normalisation is more appropriate:
\begin{eqnarray}
\sigma_{ij}=\frac{\sigma}{d_i d_j}.
\label{GeneralNorm}
\end{eqnarray}
This reflects that in any interaction
there are two factors:
the capacity of
the agent generating information and
that of the agent digesting it.
Each factor imposes dissipation of cognitive
effort according to the number of
links supporting interactions.

For the complete
graph then
\begin{eqnarray}
\sigma^{\rm{Team}}_{ij}= \frac{\sigma}{(N-1)^2}.
\end{eqnarray}
Contrasting an all-to-all team, we consider a star graph of size $N$ as paradigmatic of
a centralised leader over
$N-1$ subordinates,
where
\begin{eqnarray}
\sigma^{\rm{Lead}}_{ij}= \frac{\sigma}{(N-1)}.
\end{eqnarray}

We use frequencies from the range $[-1,1]$ using three distribution choices:
grid spacing of $N$ frequencies
within the range; randomly chosen
from a uniform distribution
in the range; and truncated
Gaussian distributed frequencies
with the variance of the distribution inside the range.
We numerically solve for a range of $N$.

As usual, we 
measure the synchronisation through the order parameter $r$
using 
\begin{eqnarray}
r e^{i\psi} = \frac{1}{N}  \sum_{j=1}^N
e^{i \theta_j} 
\end{eqnarray}
where $r$ closer to unity 
represents greater synchronisation across oscillators. 
Here, $\psi$ gives the phase of the centroid of the 
overall oscillator cluster.

\begin{figure}
\includegraphics[width=.49\textwidth]{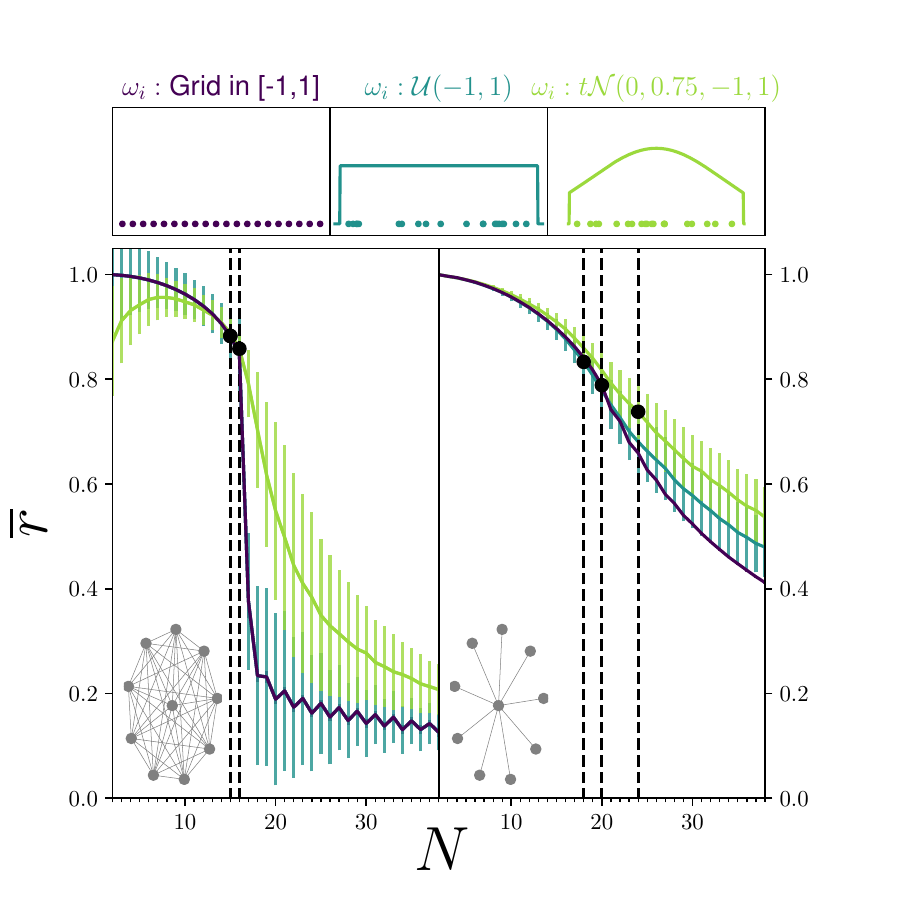}
\caption{\label{fig:order}
The average order parameter $\overline{r}$ over 1,000
realisations as a function of $N$ for three cases of frequency distributions with fixed $\sigma =20$. Examples of a sampling from these distributions are provided for $N=20$ on the upper plot. In the quadrant plots, we consider the two cases of a complete graph/"Team" (left) and a star graph/"Lead" (right). Critical values of $N$ are indicated with black dots.}
\end{figure}

We compute $r$ as a function of
$N$,
averaged over frequencies for the three distributions choices, and
over time dropping an initial transient. These are
shown in left hand panels of Fig.\ref{fig:order}.
Error bars denote 
one standard deviation over
1000 random frequency instances.

We observe
phase-transition like behaviour at a critical value of $N$. The grid frequency choice gives the sharpest `first-order' drop-off while random choices give softer transitions.
In the right hand panels of 
Fig.\ref{fig:order} we show how
the second order derivative with respect to $N$ can determine the critical point
for the softer transitions,
matching well with the 
drop in $r$ for
grid choice.

We have separately verified that increasing coupling gives the
transition at larger $N$: higher coupling supports synchronisation in larger
organisations. In the following we analytically explain
this, specifically the
frequency distribution role in the transition sharpness.

\section{Analytical treatment three ways}
We use three approaches to derive
a critical size of the complete graph
at which synchronisation collapses.
Two use clustering approximations,
appropriate for finite systems. The third modifies
Kuramoto's large $N$ approach.

The clustering approximation works
close to the break-down of team coherence,
where there are nominally two clusters
around phase values
$B_1$ and $B_2$ of 
$N_1$ and $N_2$ nodes respectively
\cite{kalloniatis2016fixed}.
Truncating the equations
to zeroeth
order (ignoring fluctuations), averaging over the
sub-structures and subtracting
the equations yields
dynamics for the phase
difference $\alpha=B_1-B_2$:
\begin{eqnarray}
\dot{\alpha} = 
\bar{\omega}_1 - \bar{\omega}_2
- \frac{\sigma(N_2-N_1)}{(N-1)^2}
\sin\alpha .
\label{twocluster}
\end{eqnarray}
Here $\bar{\omega}_i$ is
the average frequency within the
$i-$th cluster. 
The critical condition is when $\alpha$ is static;
when dynamical, the two clusters are moving
with respect to each other and
the system is not internally synchronised. Setting $\dot{\alpha}=0$,
we have an equation for
$\sin\alpha$, which
ceases to have solutions when 
$\sin\alpha=\pm 1$, giving:

\begin{eqnarray}
|\bar{\omega}_1 - \bar{\omega}_2 |
(N-1)^2
= \sigma |N_2-N_1|.
\end{eqnarray}

Contrasting with \cite{kalloniatis2016fixed}, we use this to determine the critical team size, $N_c$.

Though presented intuitively, more
rigorous treatment, including fluctuations and stability analysis, are provided in \cite{kalloniatis2016fixed};
with stochastic effects in
\cite{holder2017gaussian,kalloniatis2019two}. Specifically, for a two-cluster fixed
point analysis of the Kuramoto model, Lyapunov instability
occurs at precisely the
point where Eq.(\ref{twocluster})
has no static solution.
(The dynamics around this
critical point obey a linear ratchet potential
\cite{Linder2001,Mulhern2011,Zuparic2017}, also called
the Adler equation \cite{Adler1946}.)

We now consider the most
likely sequence to clustering.
Firstly, we label the nodes according
to the frequency order:
$-\omega_{\rm{max}}< \omega_1 < 
\omega_2 < \cdots < \omega_{N-1}
< \omega_N=\omega_{\rm{max}}$. 
With the symmetry of the complete
graph, the first oscillator to break-off the main
cluster (as $N$ increases) will
be the most distant from the mean
$\bar{\omega}=0$. 
Thus $N_1=N-1, N_2=1$. Moreover,
$\bar{\omega}_1\approx 0$ since
it involves the oscillator-bulk, so
$\bar{\omega}_2\approx \omega_{\rm{max}}$.
Thus we arrive at
\begin{eqnarray}
N_c=\frac{1}{\omega_{\rm{max}}} \sigma +1.
\label{NcApprox}
\end{eqnarray}

An improvement
on this uses a three-cluster approach \cite{kalloniatis2016fixed}, intuitively motivated
because for a symmetric frequency distribution, the likely fragmentation
from the main group will be at the two
extremes of the oscillator set.
Adapting section 3.6 of \cite{kalloniatis2016fixed}
we let
$B_0$ denote the main cluster
phase,
and $B_1$ and $B_2$ those of the two extreme
oscillators. Thus,
$\omega_1=-\omega_{\rm{max}}$
in $B_1$ and $\omega_N=+\omega_{\rm{max}}$
in $B_2$.
Using $\alpha_{B_0B_1}$ 
as the phase between the main cluster centroid and one extreme,
and $\alpha_{B_1B_2}$ for the phase
between the two extremes
we obtain their
dynamics from truncation
of the Kuramoto equations:
\begin{eqnarray}
\dot{\alpha}_{B_0B_1}
&=& \bar{\omega}_0 - \bar{\omega}_1
- \frac{\sigma}{(N-1)} \sin\alpha_{B_0B_1} \nonumber \\
&& + \frac{\sigma}{(N-1)^2} \sin\alpha_{B_1B_2} \nonumber \\
&&- \frac{\sigma}{(N-1)^2} \sin(
\alpha_{B_0B_1} + \alpha_{B_1B_2})\\
\dot{\alpha}_{B_0B_1}
&=& \bar{\omega}_1 - \bar{\omega}_2
+ \frac{\sigma(N-2)}{(N-1)} \sin\alpha_{B_0B_1} \nonumber \\
&& - \frac{2\sigma}{(N-1)^2} \sin\alpha_{B_1B_2} \nonumber \\
&&- \frac{\sigma(N-2)}{(N-1)^2} \sin(
\alpha_{B_0B_1} + \alpha_{B_1B_2})
\end{eqnarray}
We have used here
that the degree of an extreme
node to the main cluster is $(N-2)$
and the two extreme nodes have
one link between them.
Also, $\bar{\omega}_0,\bar{\omega}_1,\bar{\omega}_2$ represent
the averages of the frequencies
in the main and extreme clusters.
Straightforwardly then
$\bar{\omega}_0\approx 0,\bar{\omega}_1=-\omega_{\rm{max}},\bar{\omega}_2=+\omega_{\rm{max}}$.
At locking, $\alpha_{B_0B_1}= \alpha_{B_1B_2}=0$.
Because of the sine of the phase angle sum, these remain
analytically intractable, even
using addition formulae.

However, we may set the extreme
case again where the extremal
oscillators are $\pi$ out of phase
and each $\pi/2$ away from
the main cluster. Thus
$\sin\alpha_{B_0B_1}=1,\sin\alpha_{B_1B_2}=0 $.
This yields (from either equation)
\begin{eqnarray}
\omega_{\rm{max}} - 
\frac{\sigma (N-2)}{(N-1)^2}=0
\end{eqnarray}
Solving for $N$ 
gives
\begin{eqnarray}
N_c= 1 + 
\frac{\sigma}{2\omega_{\rm{max}}} \pm  \frac{\sigma}{2\omega_{\rm{max}}}
\sqrt{1-\frac{4\omega_{\rm{max}}}{\sigma}}.
\end{eqnarray}
Naturally this is only valid for
$\sigma>4\omega_{\rm{max}}$.
This provides again for the same
linear dependence on coupling but
with a non-linear correction
for small $\sigma$. In fact, expanding the
square root, shows the constant
term cancelling, thus 
$N_c=\frac{\sigma}{\omega_{\rm{max}}} +
{\cal{O}}(\sigma^{-1})$.
Non-symmetric cases such as
$\bar{\omega}_1=\omega_{\rm{min}}
\neq -\omega_{\rm{max}}$
become analytically intractable.

Finally, we consider the analogue
to Kuramoto's derivation of the
critical coupling for $N\rightarrow \infty$. We follow, for
example, Strogatz's presentation
in \cite{strogatz2000kuramoto}. 
With $K=\frac{\sigma}{N-1}$
much of the derivation proceeds
{\it mutatis mutandis}. Kuramoto's equation
may be combined with the order parameter in
the centre of mass frame giving
$\dot{\theta}_i = \omega_i - K r \sin\theta_i$ so at steady state,
locked oscillators will satisfy
$|\omega_i| \leq Kr$ with 
$|\theta_i|\leq \frac{\pi}{2}$,
with the remaining oscillators drifting; a steady state distribution
of oscillators
is assumed here.
Locking will occur for those in the
frequency distribution centre.
The essence is to
compute $re^{i\psi}=r$ (in the centre
of mass frame) within the ensemble of
the steady state oscillator distribution
as a sum of locked and drifting
populations. As $N\rightarrow \infty$, 
the drifter population
contributes zero to the
ensemble average $\langle e^{i\theta}\rangle$ due to
symmetry, while the locked
group gives $\langle e^{i\theta}\rangle=
\langle \cos\theta \rangle$,
re-expressible as an integral
over the frequency distribution $g(\omega)$.
Changing variables, this
gives 
\begin{eqnarray}
r=Kr \int_{-\frac{\pi}{2}}^{+\frac{\pi}{2}} d\theta \cos^2\theta g(Kr\sin\theta),
\end{eqnarray}
a self-consistent
nonlinear equation for $r$. The non-trivial branch of this
bifurcates as $r\rightarrow 0$
at $K_c = \frac{2}{\pi g(0)}$.
However, now re-expressing $K_c=\sigma/N_c$
yields
\begin{eqnarray}
N_c = \frac{\pi g(0)}{2} \sigma.
\label{KurCritN}
\end{eqnarray}

Given the assumed finiteness
of $K$ here, a 
linear dependence of
$N_c$ on $\sigma$ was inevitable.
However we have now determined the
coefficient. For a rectangular
normalised distribution $g$ of width $2\omega_{\rm{max}}$ we have
$g(0)=\frac{1}{2\omega_{\rm{max}}}$.
We thus obtain
$N_c=\frac{\pi}{4} 
\frac{\sigma}{\omega_{\rm{max}}}$.
Compared to the clustering approach, the coefficient
``1'' is replaced by $\frac{\pi}{4}\approx 0.785...$,
a consequence of replacing
a point mass (cluster) by a uniform
spread of oscillators.

We give all three of these
derivations because for intermediate
$N$, at sizes the literature
suggests teams lose coordination,
the strict assumptions 
behind any one approach may
break down.
For example, with evenly-spaced frequencies in 
$[-\omega_{\rm{max}},\omega_{\rm{max}}]$
the system is biased towards
the symmetric distribution
assumed in the $N\rightarrow\infty$
arguments. Then 
the slope based on Kuramoto's
result may be expected.
Contrastingly, for any 
instance of random 
selections of frequencies
symmetry will not be
exact. A two cluster configuration
may then be reasonable, an extreme
oscillator imbalancing the distribution.

\begin{figure}
\includegraphics[width=.49\textwidth]{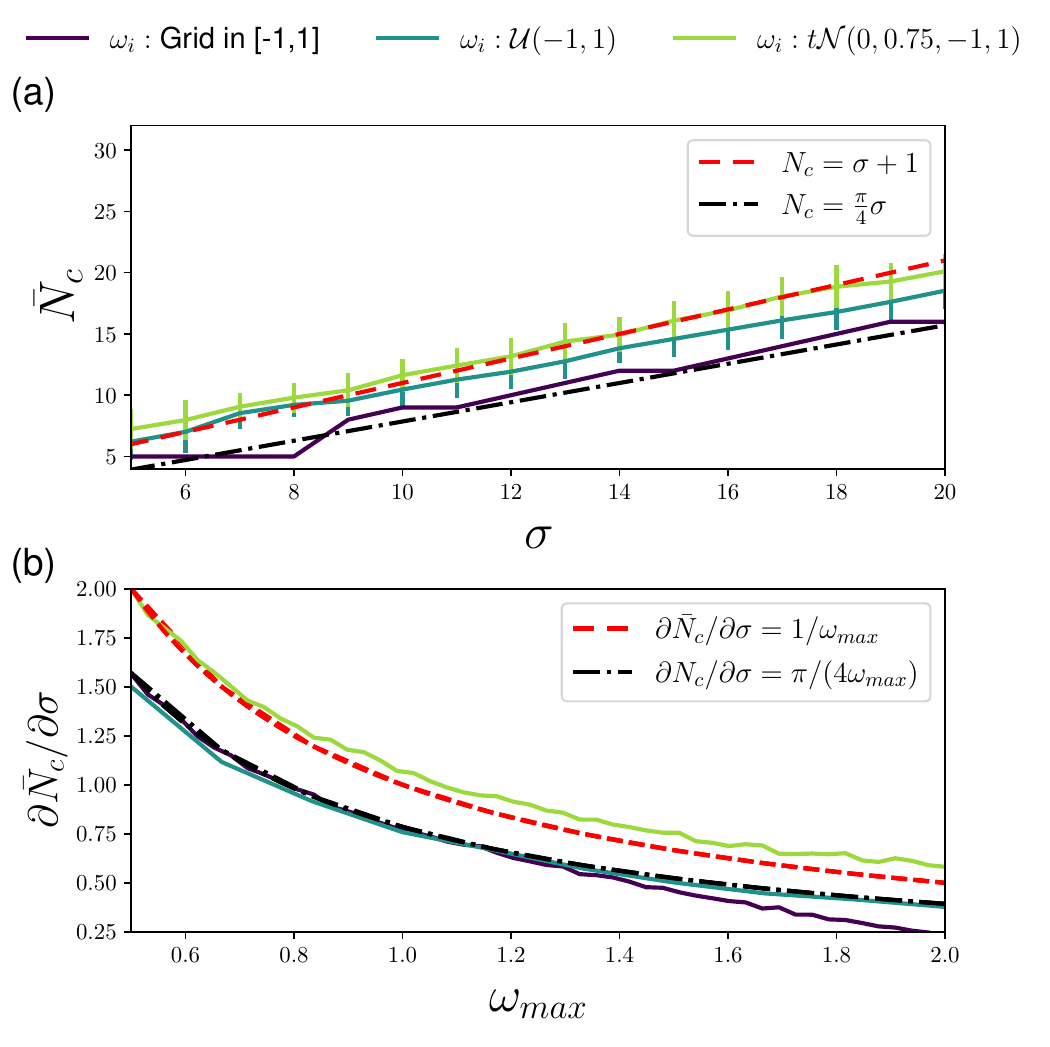}
\caption{\label{fig:Nc}
Plots of the critical size $N_c$
vs. $\sigma$ (top) and the slope of $N_c$ vs. $\omega_{\rm{max}}$ (bottom) for three different frequency distributions. Dashed lines represent analytical 
dependencies given in the figure legends.
}
\end{figure}

In Fig.\ref{fig:Nc}(a) we plot
the critical size $N_c$ 
(using the second order derivative) as a function
of coupling for the different
frequency distribution choices
where we have set the maximum frequency to one, effectively normalising. Superimposed
are the analytical results 
Eq.(\ref{NcApprox}) and Eq.(\ref{KurCritN}).
We see that both
results are valid for different
frequency choices. The grid choice, as suggested by replicating the symmetry realised for
a large number of oscillators, 
shows a slope following
Kuramoto's result.
On the other hand, the truncated
Gaussian, by creating a cluster of
frequencies in the centre, leads to approximately unit-value
slope.
In panel (b) we vary the
maximum frequency and verify
the $1/\omega_{\rm{max}}$ 
dependence of the slope.

In summary we find that the critical
team size is
\begin{eqnarray}
N^{Team}_c = \xi \frac{\sigma}{\omega_{\rm{max}}}
\label{Nc-final}
\end{eqnarray}
with $\xi$ a characteristic constant reflecting the clustering
of the main group of oscillators
close to fragmentation.

    \label{fig:my_label}

\section{Calibrating coupling for teams}
We argue that compact frequency
distributions are appropriate for teams.
With $\omega$ representing
an intrinsic decision-making speed
in any competent `professional' team,
there will be some lower bound
representing limitations
on admission to team-membership through
recruitment, training, and promotion. The upper bound represents the biological
limitation of human information processing times. This is natural for rational deliberate
decision-making, or `system two'
in Kahneman's classification \cite{kahneman2011thinking}. But cognitive processing time is
finite, though faster, 
even for intuitive `system one'
decisions \cite{morewedge2010associative}, 
which are still conscious acts
\cite{Graf2018}, often described using
Klein's Recognition Primed model
\cite{klein2017sources}.

We may thus compute $\sigma/\omega_{\rm{max}}$ using the complete
graph results using the empirical bounds
$5 \leq N_c \leq 8$.
With Eq.(\ref{Nc-final})
we obtain the calibrated coupling:
\begin{eqnarray}
\frac{5}{\xi} \leq 
\frac{\sigma}{\omega_{\rm{max}}}
\leq
\frac{8}{\xi} .
\end{eqnarray}

\section{Validation: Span of Control}

\begin{figure}
    \centering
    \includegraphics[width=.49\textwidth]{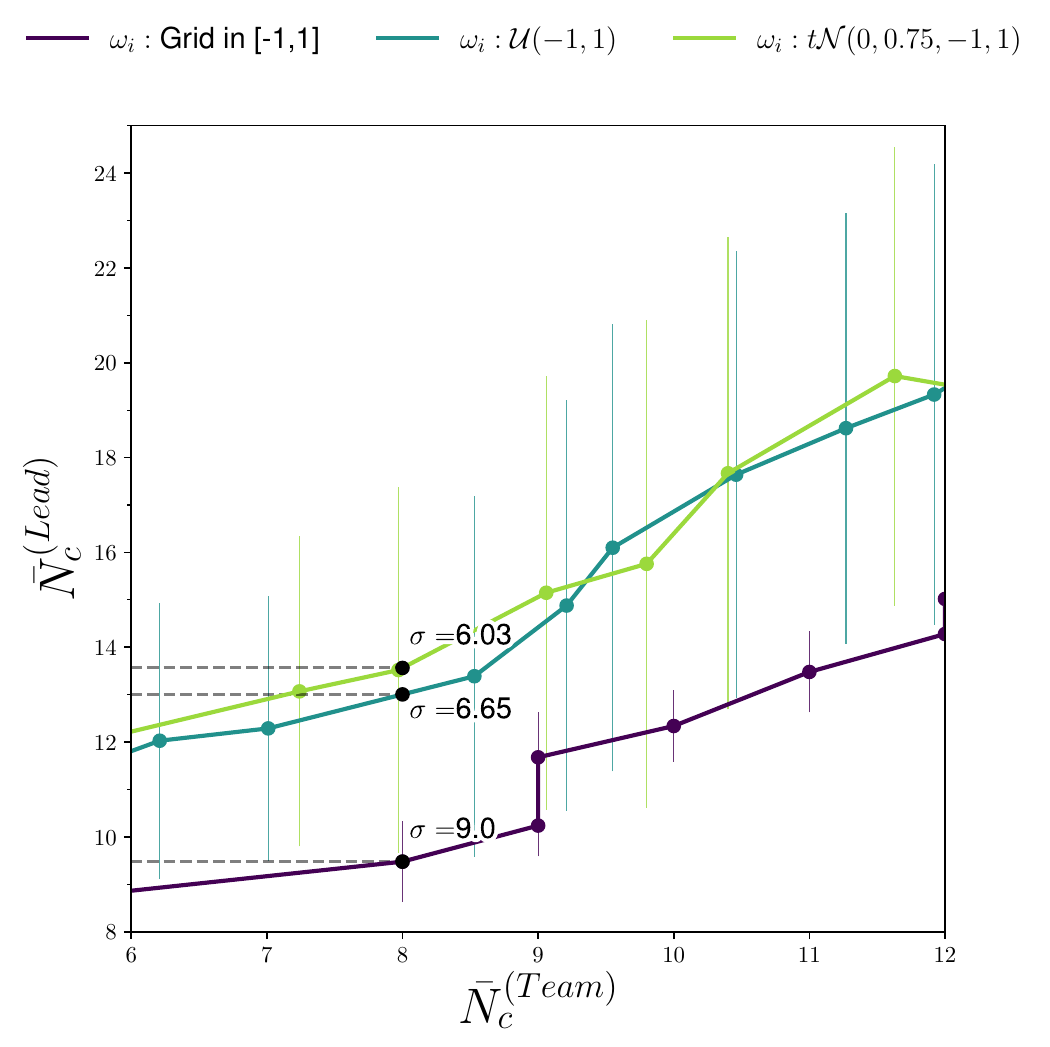}
    \caption{Relationship between the mean value $\bar{N}_c$ for the team and leader cases for equivalent values of $\sigma$ over 100 samples from the three frequency distributions. Bars represent the standard deviation and black dots highlight the values of $\sigma$ when $\bar{N}_c^{(Team)} =8$.}
    \label{fig:Critcal teams vs leaders}
\end{figure}

With coupling calibrated we test against the leader/supervisor 
organisation, modelled
as the star graph.
Here the maximum 
number of subordinates a supervisor can oversee is known as
the span of control. 
We write the leader model
in terms of the maximum team size
by eliminating the coupling using Eq.(\ref{Nc-final}),
namely $\sigma=\omega_{\rm{max}}N^{Team}_c/\xi$. 
For the three different frequency distributions we may
compute the critical leader organisational size $N_c^{Lead}$ in terms
of the team size for fixed $\xi$,
shown in Fig.~\ref{fig:Critcal teams vs leaders}. Note that
the grid frequencies give the narrowest error and lowest $N_c^{Lead}$ while requiring higher coupling.
Taking the sample averages from the studied frequency distributions, we find that $9<N^{Lead}_c< 14$
given a calibration of coupling to a team structure, for example $N_c^{Team}=8$.

The literature for $N_c^{Lead}$ shows significant divergence,
as underscored by Mintzberg \cite{Mintzberg1979}. Data here comes not from controlled studies but {\it in vivo} industrial or business
firms with all their inherent complexity. Management theorist, Urwick \cite{Urwick1956} proposed
five, maximum six, as the 
maximum span of control.
Numerous twentieth century studies
found a median of ten, with a maximum
of 14. Underlying this variation
are the many factors present in real world management: technology support systems at different epochs, supervisor managerial style (micro-managing vs delegation), and 
whether monitoring of the cross-interactions
of the subordinates and interactions where the
leader intervenes between multiple subordinates also impose
on supervisor cognitive-capacity.
For the latter case, the
formulation of Graicunas \cite{Graicunas1937}
is significant, proposing
a maximum of six. However, this
construct is quite different from
our formulation
as a star graph,
a particular instantiation of
the span of control problem.
Our result with larger $N_c^{Lead}$ than Graicunas
is consistent with a leader only managing subordinates and not `micro-managing' their interactions. It is also consistent
with the maximum size of 14 in later studies which generally abjure
such micro-management.


\section{Conclusions}
We have adapted the Kuramoto model
to represent human organisations.
Our analytical expressions
for the critical team size are consistent with numerical calculations. Using data
on viable team size we calibrated
the coupling. This then
determined a range for the maximum
span of control. The literature,
often using only combinatorial arguments, 
is imprecise on this value,
but our result overlaps with
its recommendations.
Beyond \cite{HouseMiner1969},
few works analyse the two different
organisational constructs within
one quantitative approach.

By transcending combinatorics and incorporating the
decision-making dynamics, our formulation models a richer
variety of organisational forms, offering experimentation opportunities for quantitative
organisational research. For example, are the maximum team size and span of control
the same to short time-scales?
Can this approach be integrated
with representations of technological artefacts?

In this spirit, our formulation bridges traditional technological uses of the Kuramoto model with future socio-technical applications 
for human-machine teams as the current Artificial
Intelligence revolution impacts society.

\end{document}